\documentclass[11pt]{article}

\usepackage[margin=1.25in]{geometry}
\usepackage{authblk}
\usepackage{mathptmx}       
\usepackage{helvet}         
\usepackage{courier}        
\usepackage{type1cm}        
\usepackage{makeidx}         
\usepackage{graphicx}        
\usepackage{multicol}        
\usepackage[bottom]{footmisc}

\usepackage{graphicx}
\usepackage{amsmath, amssymb, mathrsfs}
\usepackage{algorithm}
\usepackage[noend]{algpseudocode}
\usepackage{xcolor}
\usepackage[numbers]{natbib}

\begin{document}
\title{Exploring the Role of Intrinsic Nodal Activation on the Spread of Influence in Complex Networks}
\date{}

\author[$\dagger$]{Arun V. Sathanur}
\author[$\dagger$]{Mahantesh Halappanavar}
\author[$\star$]{Yi Shi}
\author[$\star$]{Yalin Sagduyu}

\affil[$\dagger$]{Pacific Northwest National Lab, Richland, Washington 99352}
\affil[$\star$]{Intelligent Automation, Inc., Rockville, MD 20855}


%
%
\maketitle

\abstract{In many complex networked systems, such as online social networks, activity originates at certain nodes and subsequently spreads on the network through influence. In this work, we consider the problem of modeling the spread of influence and the identification of influential entities in a complex network when nodal activation can happen via two different mechanisms. The first mechanism of activation stems from factors that are intrinsic to the node. The second mechanism comes from the influence of connected neighbors. After introducing  the model, we provide an algorithm to mine for the influential nodes in such a scenario by modifying the well-known influence maximization algorithm to work with our model that incorporates both forms of activation. Our model can be considered as a variation of the independent cascade diffusion model. We provide small motivating examples to facilitate an intuitive understanding of the effect of including the intrinsic activation mechanism. We sketch a proof of the submodularity of the influence function under the new formulation and demonstrate the same on larger graphs. Based on the model, we explain how influential content creators can drive engagement on social media platforms. Using additional experiments on a Twitter dataset, we then show how the formulation can be applied to real-world social media datasets.  Finally, we derive a centrality metric that takes into account, both the mechanisms of activation and provides for an accurate, computationally efficient, alternate approach to the problem of identifying influencers under intrinsic activation.}


\section{Introduction and Related Work}
\label{sec:intro}

The advent and rapid adoption of social media platforms allows people to self-organize into complex social networks with rich dynamics. Users can disseminate their views, opinions, and other content while simultaneously consuming and reacting to the content created by the friends, people, and organizations they follow. The success of such platforms depends on the myriad of content creators, the quality of their content, and the activities their audiences generate because of the various types of engagement possible with the posted content. These actions can be attributed to influence. The dynamics of influence and resulting diffusion of information in complex networks has been the subject of intense scrutiny for researchers and practitioners in many fields with particular attention to the identification of central or influential nodes on the network. One rigorous approach to finding influential users with motivations originating in viral marketing is the approach based on {\em influence maximization}. 

We can define the influence maximization problem as follows : Consider a directed graph $G=(V,E)$ that abstracts a complex network, where $V$ is the set of nodes $V  = \{v_1,v_2,v_3\dots\}$ and $E$ is the set of directed edges  $\{\left(v_u,v_w\right) | v_u$, $v_w \in V \}$.  The directed edge $\left(v_u,v_w\right)$ implies that $v_u$ can influence $v_w$ and not the other way round. However, it is possible that both $\left(v_u,v_w\right)$  and $\left(v_w,v_u\right)$ are valid edges. We denote by $|V|$ the total number of nodes and by $|E|$ the total number of edges in the graph $G$. Further, the nodes are labeled as either \textit{Passive} or \textit{Active}, denoting the state of the vertex. A necessary but not sufficient condition for an active vertex $v_u$ to activate a passive vertex $v_w$ is that $\left(v_u,v_w\right) \in E$. Other conditions come from the nature of the diffusion model. Given that it is possible to initially activate $k$  nodes, the influence maximization problem aims to find the particular set of $k$ \textit{seed nodes}, called the \textit {seed set} $S$. When the nodes in the set $S$ are activated, the spread of influence results in maximal activations on the network among all possible such sets of $k$ nodes. Note that in the subsequent discussions, we use the terms \textit{reachability}, \textit{number of activations on the network}, and \textit{influence spread} synonymously to denote the total number of active nodes on the network after running the diffusion models, starting from the initial set of active nodes, until no more activations are possible.
	
	Starting with the landmark paper by Kempe, Kleinberg, and Tardos \cite{Kempe:2003}, several works have explored newer diffusion models and variations to the ones studied in the work by Kempe et al., namely, the independent cascade (IC) model and the linear threshold (LT) model. These models explicitly address the various sociological aspects of influence.  Li {\em et al.} in \cite{li2013influence} consider influence dynamics and influence maximization under a general voter model with positive and negative edges. In a follow-up work, Kempe et al. \cite{kempe2005influential} discuss a diverse set of models including the so called decreasing cascade model where attempts by multiple neighbors to activate a node results in decreasing probability of activation, as the size of the set of neighbors trying to activate the node increases. The authors in ref. \cite{srivastava2014influence} propose a general diffusion model that takes into account different granularities of influence, namely pair-wise, local neighborhood etc. The authors in \cite{chen2011influence}, consider influence maximization under the scenario where negative opinions may emerge and propagate. In  \cite{gionis2013opinion}, the authors consider the problem of identifying the individuals whose strong positive opinion about a product will maximize the overall positive opinion about the product. In the process, the authors leverage the social influence model proposed by Friedkin and Johnson \cite{friedkin1999social}. For a comprehensive survey on the various models of influence, we refer the reader to the paper by Zhang et al. \cite{zhang2014recent}.

Next, we consider the models that address two different types of activation : intrinsic and influenced.  The interplay of these two mechanisms are exemplified by three different scenarios outlines below. 
\begin{itemize}
\item Users posting content on social media due to their own initiative constitutes intrinsic activation. Actions such as sharing , re-tweeting , commenting constitute influenced activation. 
\item Posting behavior that is external to a given network can be considered to be intrinsic activation. This would include watching a video on a website from a shared email link and then sharing it on Twitter. From the perspective of just the Twitter network, it appears that such users are intrinsically activated. 
\item In a traditional social network, such as a physical community, that is not an OSN (online social network), intrinsically activated users would be those who take the initiative to start an activity, for example a campaign for social good. The same can then spread through word of mouth, flyers etc.
\end{itemize}

Myers, Zhu, and Leskovec investigate the diffusion of information, with origins external to that of a social network, through the internal social influence mechanism \cite{myers2012information}.  In a recent work \cite{farajtabar2014shaping}, the authors recognize that the events on social media can be categorized as exogenous and endogenous and model the overall diffusion through a multivariate Hawke's process  to address activity shaping in social networks. In another recent work, the authors in \cite{Quach2016} propose a novel diffusion model based on factor graphs and graphical models where the node potentials can correspond to the notion of intrinsic activation in our case. However, the focus of their work in on the diffusion model itself, not on the aspects of intrinsic activation. While being similar in spirit to these works, our work is geared toward modeling the spread of influence and mining influential nodes in scenarios with intrinsic and influenced activation - aspects that have not been studied in existing literature.
	
We make the following contributions in this work. 

\begin{enumerate}
\item  Our approach results in a probabilistic model for two different types of nodal activations, namely intrinsic and influenced mechanisms found in real-world networked systems, such as OSNs. 
\item We examine these mechanisms in the context of influencer mining from \textit{two different perspectives}: the well-known combinatorial influence maximization perspective and a generalized centrality perspective. 
\item We define a modified influence spread function, sketch a proof of its sub-modularity, and provide a modified version of the influence maximization algorithm to maximize the new influence spread function
\item We examine the nature of content creators and consumers on a social network in light of the two activation mechanisms.
\item Carefully chosen experiments on synthetic and real-world graphs are used to illustrate various aspects of the model and compare it to the independent cascade model.
\item We derive a new centrality metric from the activation model and show that this metric can accurately identify influential users in a computationally efficient manner. 

\end{enumerate}

The initial aspects of this work was published in \cite{Sathanur2016}. The present version is a significant extension of the above work where we have extensively examined the content creation and content spreading mechanisms, formally sketched a proof of the submodularity of the modified influence function, added an extensive set of experiments on a real-world Twitter graph and improved the overall narrative by means of several smaller additions. 

\section{Modified Influence Maximization Approach}
\label{sec:modification}

\subsection{Formulation}
\label{sec:formulation}
Considering that nodal activation can originate from two different mechanisms, \textit{Intrinsic} and \textit{Influenced}, allows us to effectively model the so-called \textit{self-evolving} systems (eg. OSNs) that are comprised of content creators (higher probability of getting activated intrinsically) and content spreaders (activated via  influence and spreading the information). Recognizing that most of the users are in some sense both activity creators and content spreaders (typically also the content consumers) at the same time, we introduce a real-valued parameter $\alpha \in [0,1]$ that models the probability of self activation. The total probability for activation of a given node (user) $i$ is composed of the probability of activation from the two different mechanisms. The parameter $\alpha(i)$ denotes the probability of intrinsic activation, and $\beta(i)$ denotes the probability through influence with $\alpha(i) \ge 0$ and $\beta(i) \ge 0$. The influenced part of the probability for activation is comprised of the activation probabilities due the 1-hop neighbors of the user under consideration. 

\begin{figure}[htbp]
\begin{center}
\includegraphics[width=10cm]{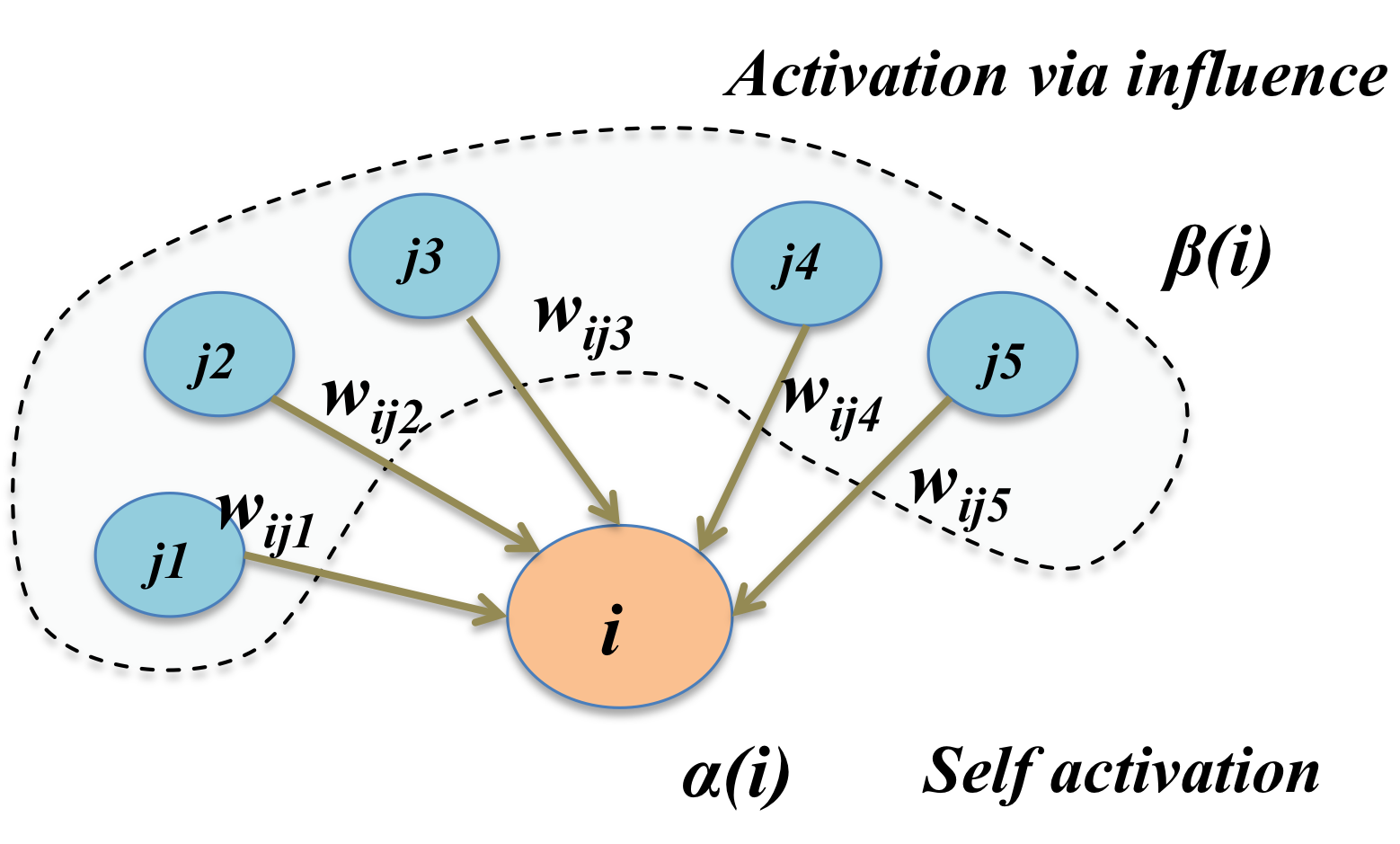}
\caption{A concise representation of the self and influenced mechanisms of activation of a node $i$.}
\label{fig:intro}
\end{center}
\end{figure}

 Note that there are many interaction models that are studied under influence maximization as pointed out in Section \ref{sec:intro}. Our model with intrinsic activation is based on one of the most widely-studied models namely, the independent cascade model and this will be the focus of the current work. Specifically, in this work, we do not consider developing the variants of other models incorporating intrinsic activation. Similar to the IC model, the weights $w_{ij}$ ($0 \le w_{ij} \le 1$) when multiplied by $\beta(i)$ denote the probability of user $j$ activating user $i$, given that user $j$ is activated by either of the above means. Figure \ref{fig:intro} describes these mechanisms and the associated coefficients. The described probabilistic formulation has similarities to the Friedkin-Johnson social influence model for opinion change \cite{friedkin1999social} where the authors recognize that the dynamics of opinion change are governed by two mechanisms - the intrinsic opinion and the influenced opinion. 

We also adopt the weighted-cascade version of the IC model by normalizing the edge probabilities \cite{Kempe:2003}, so that the expected number of nodes influencing a given node is 1. Henceforth, in this work, when we refer to the IC model, we imply the weighted cascade version of the IC model. However this is not a limitation of the model since our model can also be used in the pure IC model setting.  Thus, if $\boldsymbol{W}$ denotes the sparse weight matrix that characterizes the IC edge probabilities, we require that $\boldsymbol{W}$ be row stochastic. That is, $\sum_{j,(j,i) \in E} w_{ij} = 1 $.  Further by assuming that the nodes are not \textit{lazy} and are activated by either of the two mechanisms that we outline, we set $\beta(i) = \left(1-\alpha(i)\right)$. This will render the overall IC probability between nodes $j$ and $i$ to be $(1-\alpha(i))w_{ij}$.  While the changed probabilities denote a departure from the weighted-cascade model, when the effect of intrinsic activation is added back, the expected number of nodes activating a given node is still 1. 

Note that all the model parameters discussed can be efficiently determined either by a maximum-likelihood-based approach (as in this work) or by alternative methods such as the expectation-maximization (EM) approach followed in reference \cite{saito2008prediction}. For example, the proportion of tweets by a user $i$ that are intrinsic in nature can quantify $\alpha(i)$, while a particular weight $w_{ij}$ can be determined by the proportion of user $i$'s retweets (or influenced activity) having their origin in the activity of user $j$ that user $i$ follows. While these \textit{local influence models} can be determined in alternate ways, our goal is to find the overall influencers once these model parameters are estimated. 

Our formulation addresses the problem of identifying influential nodes on a network without explicit seeding. The original influence maximization approach with roots in viral marketing explicitly activates the seed nodes while in our formulation, the system is self-evolving in that nodes get activated intrinsically with a probability (content creation) and subsequently these activations spread (content consumption and spreading) on the network. This is the focus of the next section which describes the modifications to the original influence maximization algorithm necessary to identify the influencers under intrinsic activation.

\subsection{Algorithm for mining influential nodes under intrinsic activation}
\label{sec:algo}

We propose a simple modification to the classic influence maximization framework using the greedy hill-climbing optimizer \cite{Kempe:2003}, working with the IC model, to incorporate the self-activation mechanism. Let us assume that we are seeking $k$ influential nodes out of a total of $N$ nodes on the network. Let $S^p$ be the set of influential nodes at step $p \le k$. The greedy hill-climbing optimizer expands the set to size $(p+1)$ by polling each of the nodes not in $S^p$ and augmenting those nodes, one at a time to form the set  $S^p \cup \{v\}$ and looking for the best marginal gain in terms of the activations. At each such step $p$, instead of setting each of the nodes in $S^p \cup \{v\}$ to be activated and then computing the activations according to the IC model, we probabilistically activate each node in $S^p \cup \{v\}$ with a probability given by the corresponding $\alpha$ values to simulate the intrinsic activation process. This modification is depicted in line 9 of  Algorithm \ref{algo:fj-ic}. Given the probabilistic nature of the algorithms, the overall activation numbers are obtained by running the diffusion model in a Monte Carlo fashion by invoking $n$ independent trials involving randomized graphs with corresponding edge weights. 

\begin{algorithm}[!htb]
\begin{algorithmic}[1]
\Procedure{IC-Int}{$G, P, \alpha, k, n$}
	\State Generate $n$ random numbers $r\sp{1}\sb{uv}$ ...$r\sp{n}\sb{uv}$ for each 
	       edge in $E$ and generate a set $SG$ containing $n$ subgraphs such that in 
	       subgraph $i$, $w\sb{ij}\geq r\sp{i}\sb{uv}$ 
	\State $S \gets \emptyset$ \Comment{Set of influential nodes to be mined}
	\While{$|S|< k$}
		\State $v\sb{best} \gets \emptyset$, $a\sb{best} \gets 0$
		\For {each node $v$ in $V\setminus S$}
				\State $a \gets 0$
				\For {each $G_i \in SG$ in {\tt parallel}}
					\State $\hat{S} \gets $ active nodes in $S \cup \{v\}$ based on $\alpha$
					\State Compute number of nodes, $\hat{a}$, in $V\setminus \hat{S}$ 
				    	   that are reachable from the $\hat{S} $  
					\State $a \gets a + \hat{a}$ \Comment{Synchronized update}
				\EndFor
				\If{$a\geq a\sb{best}$}
					\State $v\sb{best} \gets v$
					\State $a\sb{best} \gets a$
				\EndIf
		\EndFor
		\If{$v\sb{best} \neq \emptyset$}
			\State $S \gets S \cup \{v\sb{best}\}$
		\EndIf
	\EndWhile
	\State \textbf{return} $S$
\EndProcedure
\end{algorithmic}
\caption{\label{algo:fj-ic} Selects a set of k influential nodes that cause maximal activations on a network, following the independent cascade (IC) model with self-activation (IC-Int). 
The inputs are a directed graph ($G=(V,E)$), set of edge probabilities ($P=\{w\sb{ij}: (ji)\in E\}$), vector of alpha values ($\alpha=\{\alpha\sb{v}: v\in V\}$), number of samples ($n$), and number of influential nodes to be identified ($k$).}
\end{algorithm}

Note that this algorithm results in the computation of a modified influence spread objective function $\sigma(S)$ (same as $a_{best}$ in the algorithm), which gives us the total number of activations on the network attributable to the multi-hop influence of nodes in the set $S$ when the corresponding nodes are activated intrinsically, in accordance with their $\alpha$ values. Thus, during this process, at the step denoted by line 10,  the nodes in the set $\left(V \setminus {S^p \cup \{v\}}\right)$ are not activated intrinsically, instead they are activated via influence. These aspects are discussed further in  Sections \ref{sec:local_influ} and \ref{sec:submodularity}. 

 The running time of Algorithm \ref{algo:fj-ic} depends on the $\alpha$ values since they affect whether a particular node is active in the given sample or not (line 9 of the algorithm). If the node is not active, then reachability will not be computed from that node. The worst-case complexity of  the algorithm is the same as that for the independent cascade model and can be derived to be $O(nk^2|V||E|)$ where $n$ is the number of Monte Carlo samples, $k$ is the number of influential nodes sought, $\left|V\right|$ is the number of nodes (vertices) in the graph and $\left|E\right|$ is the number of edges in the graph. The approach to solve the problem as detailed in Algorithm \ref{algo:fj-ic}, is based on the classic greedy algorithm to maximize monotone sub-modular functions. There are two ways to make the algorithm scalable. One is to accelerate the outer greedy optimization loop and the second method is to improve the scalability of the reachability computation. There is  prior work on both the areas. For example, the work presented in \cite{leskovec2007cost} uses lazy evaluations to speed up the greedy algorithm while reference \cite{mirzasoleiman2015lazier} uses a stochastic version of the greedy algorithm to improve the scalability. References \cite{Cohen2014} and \cite{Borgs2014} on the other hand use techniques to speed up the reachability evaluations on the sampled sub-graphs. Because our objective function is also monotone sub-modular (more on this in Section \ref{sec:submodularity}) and uses reachability computations, it can benefit from these algorithms to scale to networks with millions of nodes.

\subsection{Content creators and engagement in online social networks}
\label{sec:local_influ}

\begin{figure}[htbp]
\begin{center}
\includegraphics[width=10cm]{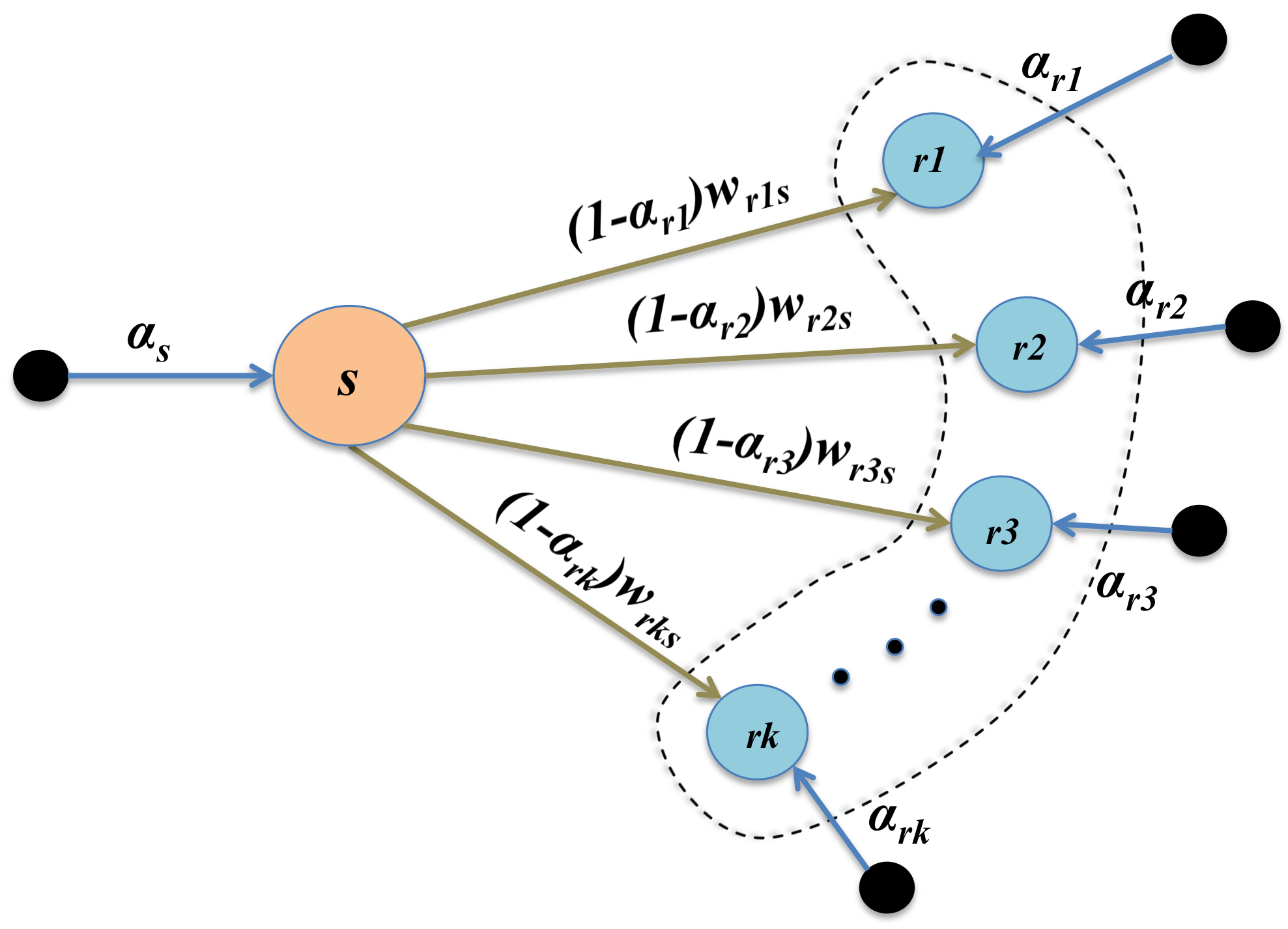}
\caption{The out-links from a source node ($s$) to the receiver nodes ($r_1...r_k$), and the associated node and edge attributes.}
\label{fig:local_outneighborhood}
\end{center}
\end{figure}

With the help of the described activation model, we examine aspects of content creation, consumption, and content spreading in OSNs, and how these are tied to the success of the platform as a whole. Figure \ref{fig:local_outneighborhood} shows the out-links around a source node  ($s$) and the various receiver (follower) nodes ($r_1...r_k$) with the $\alpha$ and $w$ values. 

Using the described activation model, it is evident that by restricting ourselves to one hop, we can write the modified influence function $\sigma(s)$ that denotes the expected number of nodes activated by node $s$ as follows:

\begin{equation}
\sigma(s) = \alpha_{s}\sum_{k=1}^{d_s^{o}}\left(1-\alpha_{r_k}\right)w_{r_{k}s}
\label{eqn:influSpreadSmall1}
\end{equation}

Here, ${d_s^{o}}$ denotes the out-degree of the node $s$. We are only modeling the activation through influence when the source node $s$ gets activated intrinsically. We do not add the term $\sum_{k=1}^{d_s^{o}}\alpha_{r_k}$ that denotes the intrinsic activation of the nodes ($r_1...r_k$). This is because, activation via influence creates engagement on the social network platform (for example sharing, commenting, liking). Thus $\sigma(s)$ can be viewed as a surrogate for engagement. The set of influential nodes (users) that we wish to compute by following Algorithm \ref{algo:fj-ic} denotes the set of influential content creators that are able to maximize this engagement (by which we mean the influence spread under intrinsic activation of the creator nodes), and are quite valuable to the platforms. 

The scenario in which all nodes have large values of $\alpha$, will result in nodes creating high volume of content on their own, and there is not much spreading of the content through different forms of engagement. Alternatively, all nodes having rather small values of $\alpha$ means that while the nodes are eager to spread the content, there is not much content created in the first place, again reducing the engagement. Therefore, we hypothesize there is an optimal assignment of the $\alpha$ values for a given assignment of the interaction probabilities and the network topology that can maximize the spread of influence under intrinsic activation. While we provide evidence of this with experiments on a real-world Twitter dataset in Section \ref{sec:twitter_graph}, solving an actual optimization problem is beyond the scope of this work. 

Equation \ref{eqn:influSpreadSmall1} provides a quick preview of the distribution of the $\alpha$ values that can lead to maximizing this engagement. The objective function $\sigma(s)$ favors a source node with large $\alpha_s$ and high out-degree, connected to receivers with low $\alpha_{r_k}$ values who easily engage with the intrinsic activity of the source node (higher value of the IC probability along these edges and lower value of the receiver $\alpha$). In practice when users can have arbitrary $\alpha$ values, Algorithm \ref{algo:fj-ic} is able to seamlessly identify  such influential content creators by simulating the two mechanisms of activation. The same will not be possible with the independent cascade model because every node that is selected to be a part of the seed set is necessarily activated thereby over-estimating a given node's influence. 

\subsection{Optimality of the influence maximization algorithm with intrinsic activation}
\label{sec:submodularity}
For the classic influence maximization problem with the IC model, the greedy hill-climbing optimizer is shown to be optimal in the sense that it provides  ($1-\frac{1}{e}-\epsilon$)  approximation guarantee on the expected influence spread function. This is because the expected influence spread $\sigma(S)$ is a monotone submodular function \cite{krause2012submodular,Kempe:2003}. The greedy algorithm expands the seed set $S$ by the addition of nodes with highest marginal gain in terms of the number of activations. For the case of intrinsic nodal activation, we have nodes activated intrinsically, as well as through influence. Thus, it appears that an influence function defined by the total number of activations on the network is not submodular. However, given that we are only interested in the total number of activations caused by the spread of intrinsic activations ($a_{best}$ in Algorithm \ref{algo:fj-ic}), the submodularity property can be shown to remain valid.  

For each node on the network $i \in V$, we can introduce an edge pointing from a newly created dummy node $i_D$ to the actual node $i$ with an activation probability equal to $\alpha_i$. This process is illustrated in  Figure \ref{fig:local_outneighborhood}. Let $V_D$ denote the set of dummy nodes. Note that there is a one-to-one correspondence between the nodes in $V_D$ and $V$. Also, because every node $i_D$ in $V_D$ has a single outgoing edge to the corresponding node $i$ in $V$, $i_D$ cannot be activated by $i$.  On the other hand, given a large number of samples $n$, the expected number of times the edge between any pair of nodes $\left(i_D \in V_D , i \in V \right)$ is activated is $n\alpha_i$ leading to us to represent the IC probability between $i_D$ and $i$ to be $\alpha_i$. This is represented by Line 9 of Algorithm \ref{algo:fj-ic}. Thus, the original influencer mining problem can now be transformed to mining for influential nodes in the set $V_D$ under the IC model. Given that the influence (cumulative reachability) function is submodular under the IC model \cite{Kempe:2003}, the influence function in the case of intrinsic activation being present, namely $a_{best}$ in Algorithm \ref{algo:fj-ic}, is also submodular.  

\section{Synthetic Experiments}
\subsection{Small organization tree}

\begin{figure}[htbp]
\begin{center}
\includegraphics[width=\textwidth]{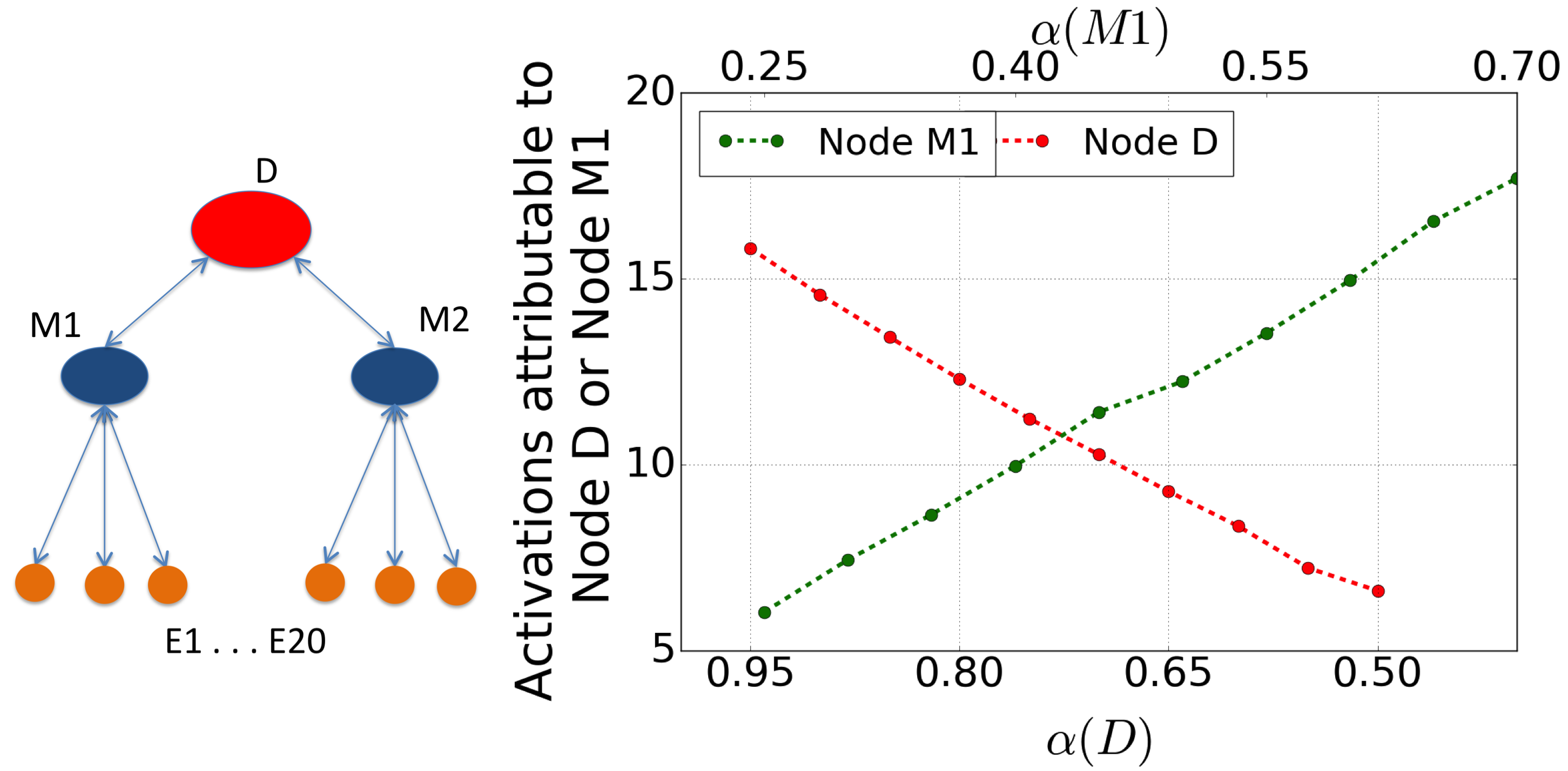}
\caption{The small organizational tree network (left) and the behavior of the influence functions with the various $\alpha$ values.}
\label{fig:small_influ}
\end{center}
\end{figure}

We first consider a small directed and weighted network with 23 nodes, organized in a tree-like fashion. The graph is depicted on the left side of Figure \ref{fig:small_influ}. In this experiment, we consider the tree-like network to depict a small organization with a Director (Node D), two Managers (M1 and M2) and 20 Employees (E1-E20), with 10 employees each working under the two Managers. We set $\alpha^{0}(D)= 0.95$, signifying that the Director almost exclusively acts intrinsically. We also set $\alpha^{0}(M1) = \alpha^{0}(M2) = 0.25$. All Employees have an $\alpha$ of 0.25 as well. As for the weights (same as the activation probabilities in the IC model), the edges ending at node D receive weights of 0.5 each (when the Director chooses to be influenced, the Director gets influenced equally by the two Managers). As for the Managers, they have a weight of 0.5 each on the edges that are incoming from D and the remaining 0.5 is split equally among the edges originating at the 10 Employees each. All Employees carry a weight of 1.0 on the edges originating from the Managers. We then perturb this baseline case to mimic a situation where the Director starts becoming more susceptible to influence, while the Manager $M1$ becomes inflexible. This is simulated by setting $\alpha(D) = \alpha^{0}(D) - \delta$ and $\alpha(M1) = \alpha^{0}(M1) + \delta$. We then sweep $\delta$ from 0.05 to 0.45. The results are shown in the right panel of Figure \ref{fig:small_influ}, where we can see that D begins as the most influential node as expected, but then M1 becomes more influential than D at a certain value of $\delta$ and will eventually have reach over most of the employees on the network. Note that the activation numbers plotted on the y-axis are the expected numbers over a Monte Carlo analysis with $n=3200$ samples. This simple experiment shows that the nature of influence spread on social networks is sensitive to the extent of intrinsic activation of the key nodes. Clearly these scenarios cannot be easily captured by the IC model, where the concept of intrinsic activation with a continuous probability value ($\alpha \in [0,1]$) does not exist. 

\subsection{Larger graphs and the influence function}
\label{sec:larger_graphs}
Our next experiment involves two larger graphs where the topology of one is from a real-world dataset, while the other is synthesized. In both cases, the node $\alpha$ values are drawn from a uniform distribution $U[0,1]$,  and the $w_{ij}$ values are also drawn from $U[0,1]$ and then normalized as described earlier in Section \ref{sec:formulation}. The graphs under consideration are described below: 
\begin{itemize}
\item LFR-1000 graph with 1000 nodes and 11433 edges is a synthetic network  that follows the generative LFR model that mimics real-world graphs \cite{lancichinetti2009benchmarks}. 
\item The PBlogs graph \cite{adamic2005political} represents a real-world blogs network and has 1095 nodes and 12597 edges. 
\end{itemize}
Further details of these graphs are discussed in \cite{Halappa:2016}. 

\begin{figure}[htbp]
\begin{center}
\includegraphics[width=10cm]{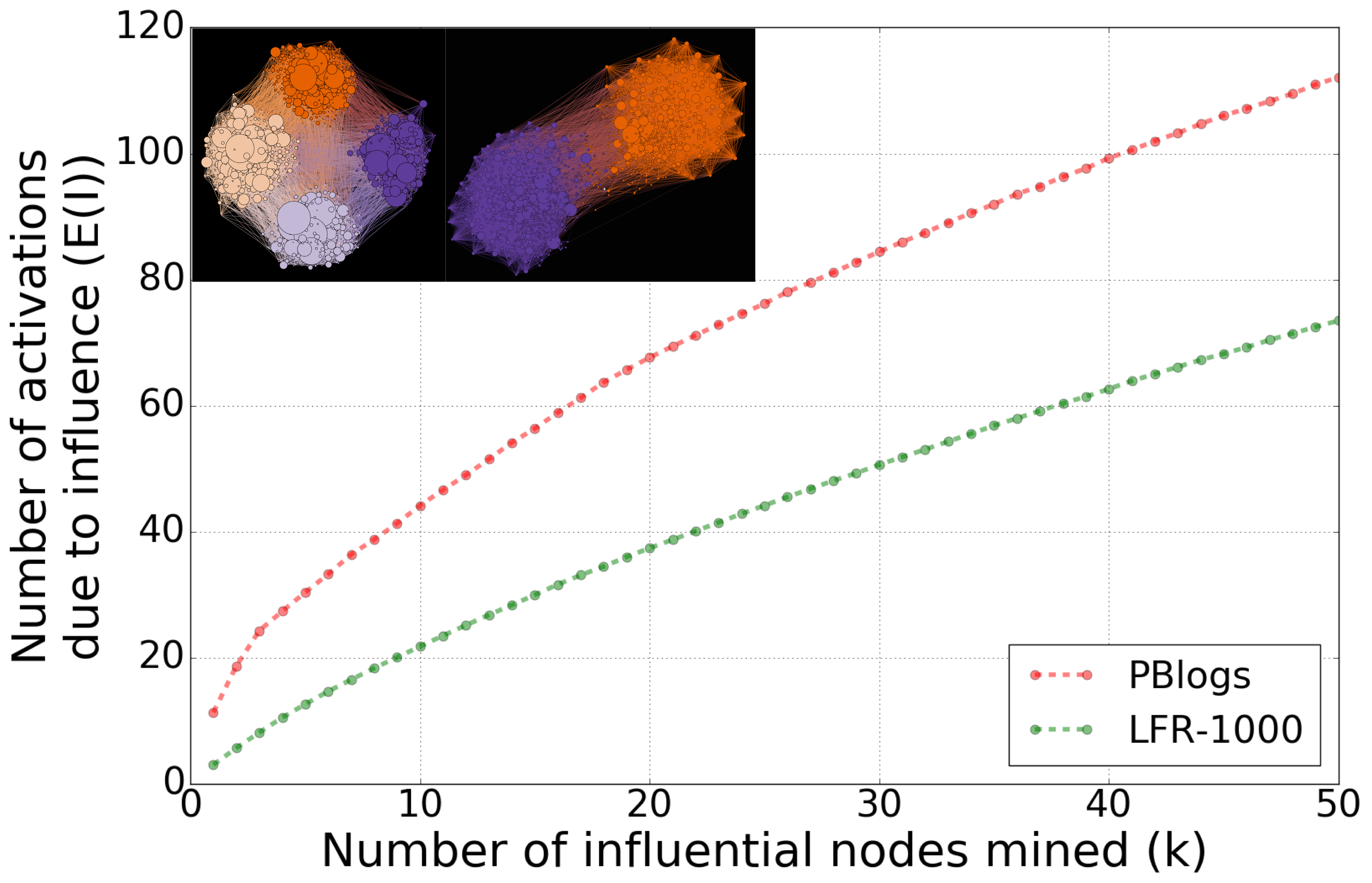}
\caption{Submodular nature of the influence function under self-activation. \textit{Inset: The PBlogs (left) and  LFR-1000 (right) networks visualized in Gephi \cite{ICWSM09154}}. The x-axis refers to the number of influential nodes mined and the y-axis refers to the expected number of activations achieved due to influence. This is represented as $E(I)$}
\label{fig:submodular}
\end{center}
\end{figure}

When we applied the modified influence maximization approach given by Algorithm \ref{algo:fj-ic} to the LFR-1000 and the PBlogs graphs and requested for 50 seeds, we observed  ( Figure \ref{fig:submodular}) that the cumulative influence spread (total number of influenced activations) showed a sub-modular character as evidenced by the diminishing gains in the total number of activations for each new seed added to the set. 


\section{Experiments on a Real-world Twitter Dataset}
\label{sec:twitter_graph}
In this section we consider the various aspects of intrinsic activation on an interaction graph constructed from Twitter data.
\subsection{Data collection}
\label{sec:data}

We first build a directed follower/friend graph from Twitter data using the public Twitter API \cite{TwitterAPI}, where each user is a vertex and a directed edge $(u,v)$ from user $u$ to user $v$ means that $v$ follows $u$.  Our goal is to capture a portion of the Twitter graph such that there are enough interactions between the nodes to estimate the $\alpha$ and the $w_{ij}$ parameter values with reasonable confidence as required by our model. Algorithm ~\ref{alg:Twitter-GraphGen} depicts the graph construction details.

 Algorithm \ref{alg:Twitter-GraphGen} starts by adding a seed twitter user $u_0$ to the set $S$ (Line $2$) and adding followers. In order to improve the density of the graph (as measured by $\frac{|E|}{|V|^{2}}$), we pick up to $k_{in}$ ($k_{in}$ is set to $15$ in our experiment) users with highest in-degree values in the set $S$ to form a new set $S^{'} \subseteq S$. If $k_{in}>|S|$, then we will just pick all the nodes in set $S$. New vertices and edges are added accordingly (Lines $9$ to $15$). Note that new vertices are added if the users are being introduced for the first time. Low out-degree nodes (based on the threshold $k_{out}$, which is set to $11$ in our experiments) are excluded in the graph construction (Line $18$). The process is repeated until required number of vertices have been added to $G$. Random seeds are added to $S$ when it becomes empty (Line $17$).

\begin{algorithm}[!htb]
\begin{algorithmic}[1]
\Procedure{Twitter-GraphGen}{$G=(V,E), k_{in}, k_{out},n$}
         \State$S = \{u_0\}$\Comment{Seed User}
         \State$V(G) \gets \{u_0\}$   
         \State$E(G) \gets \emptyset$
         \While{$|V|< n$} \Comment{Graph is less than the desired size}
                 \State$S^{'} \subseteq S$\Comment{Select $k_{in}$ users from set $S$ such as  based on top in-degree}
                 \State$S \gets S \setminus S^{'}$
                 \State$F \gets \emptyset$         
                 \For {each user $u$ in $S^{'}$}
                          \State$F \gets F \cup \{$ all followers of user $u\}$
                          \For {each node $v$ in $F$}
                                   \If{$v \notin V(G)$}
                                            \State$V \gets V \cup \{v\}$
                                            \State$S \gets S \cup \{v\}$
                                            \State$E \gets (u,v)$\Comment{Add a directed edge $(u,v)$}
                                   \EndIf
                          \EndFor
                 \EndFor
 \If{$S = \emptyset $}
 
\State$S = \{u_{rand}\}$\Comment{Randomly select a user from the dataset}
 \EndIf
         \EndWhile
         \State{Recursively remove nodes with out-degree $< k_{out}$. Goto Line 5}
         \State\textbf{return} $G$
\EndProcedure
\end{algorithmic}
 
\caption{\label{alg:Twitter-GraphGen} Generate a directed graph, $G=(V,E)$,
from the given Twitter dataset. Unique users in the dataset are represented
as nodes, and the notion of a follower is represented as a directed edge.
If user $v$ follows user $u$, then add a directed edge $(u,v)$.
The desired number of nodes $n$ and sampling parameters $k_{in}$ and $k_{out}$ are
provided as inputs to the algorithm.}
\end{algorithm}

Starting with the user ``PurdueEngineers" as the \texttt{seed\_user} $u_0$ to collect data, and following Algorithm \ref{alg:Twitter-GraphGen}, we obtain a graph with $1167$ nodes and $10292$ edges. We then generate an interaction graph from the follower/friend graph by assigning a weight on each edge $(u,v)$ based on the number of interactions, where interactions refer to replies, retweets, or mentions. We define weight $\gamma_{(u,v)}$ for a directed edge $(u,v)$ as one plus the number of times user $v$ replies to, retweets, or mentions user $u$. Note that we need to define positive weights and thus we define weight prior to normalization by one plus the number of interactions.



\subsection{Influence spread results}

Given the dataset with tweets and interactions in the form of retweets, replies, and mentions, we estimate the node-specific parameters $\alpha_i$ and the edge-specific parameters $w_{ij}$ as below. 

\begin{eqnarray*}
\gamma_i = \sum_j \gamma_{(j,i)}, \\
\alpha_i = \frac{k_i}{\gamma_i +k_i}, \\
\beta_i = 1 - \alpha_i, \\
w_{ij} = \frac{\gamma_{(j,i)}}{\gamma_i}. \; 
\label{eqn:parameters}
\end{eqnarray*}
Here, $k_i$ refers to the total number of intrinsic tweets from user $i$ and $\gamma_i$ is the total number of interactions that user $i$ participated in. Meanwhile $\gamma_{(j,i)}$ breaks this up according to the  interactions with the users that user $i$ follows. 

Note that the nature of interactions between two users can be highly complex and may be dependent on a host of features. However, in this work, we are not concerned with the complexities of the interactions. We simply compute the parameters based on counts of tweets and interactions in a {\em maximum likelihood manner} without regard to other features, such as topics and  sentiment strength.  

\begin{figure}
  \centering
  \includegraphics[width=10cm]{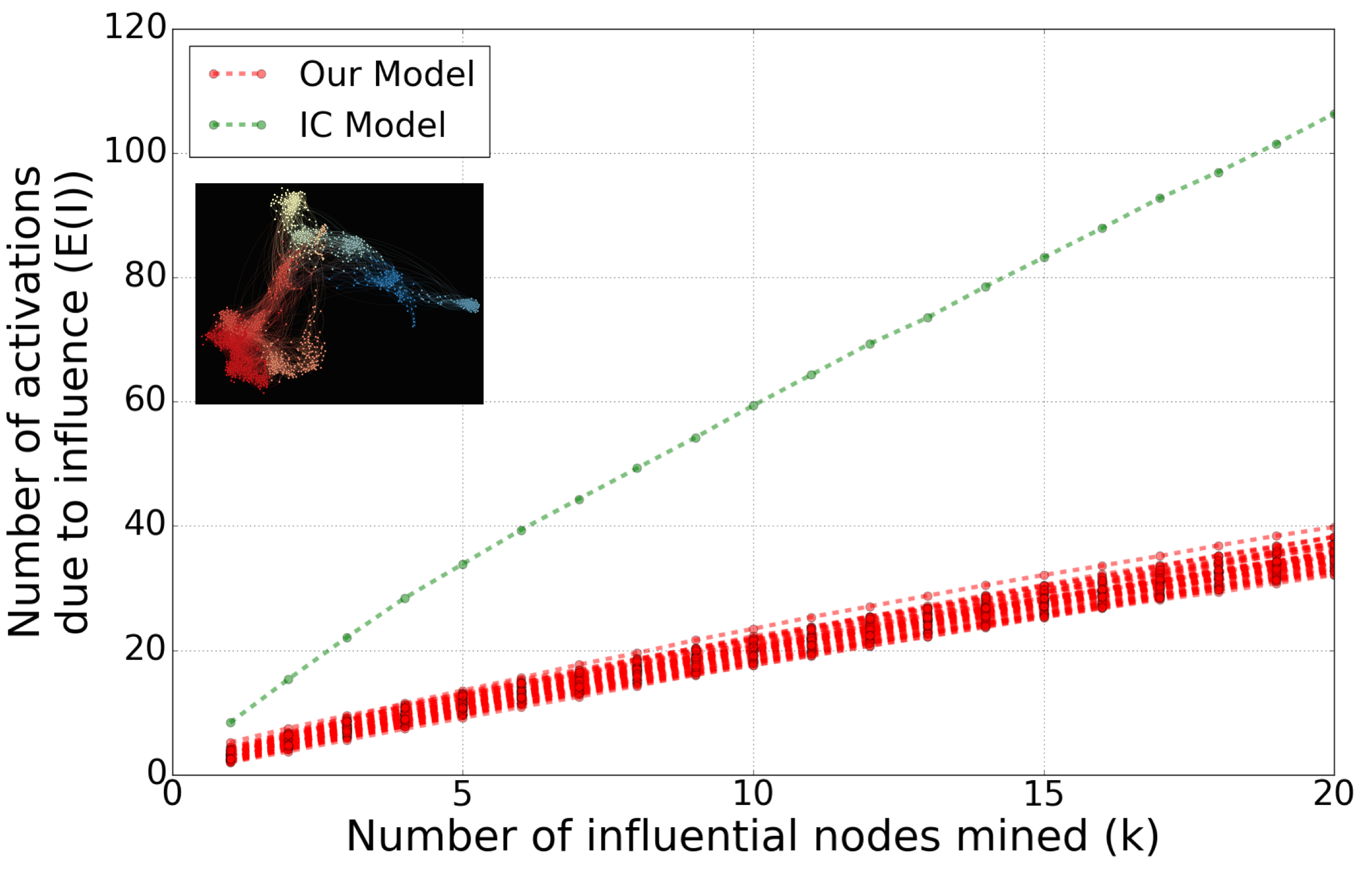}
  \caption{Comparing the influence spread with the IC model (Green) and the Monte Carlo runs on our model with intrinsic activation (Red). \textit{Inset: The constructed Twitter graph is visualized in Gephi}}
  \label{fig:cmp_twitter}
\end{figure}

On the same Twitter graph, we compare the activations achieved by the IC model and the present model, incorporating intrinsic activation. We retain the interaction probabilities as the same between the two models while noting that any interaction probability $w_{ij}$ becomes $(1-\alpha_i)w_{ij}$ for our model. We then randomized the intrinsic activation parameter $\alpha$ for each of the users to observe if the influenced activations can match that of the IC model over a number of trials. Figure \ref{fig:cmp_twitter} illustrates the results from a  Monte Carlo analysis with 50 trials.  Here the influence spread curves corresponding to the Monte Carlo runs for the model with intrinsic activation (IC-Int) are all well below the influence spread curve for the IC model as expected. Furthermore, these results are in line with the observation that the authors make in \cite{Quach2016} where they show that the IC model significantly over-estimates the activations. 

\begin{figure}
  \centering
  \includegraphics[width=10cm]{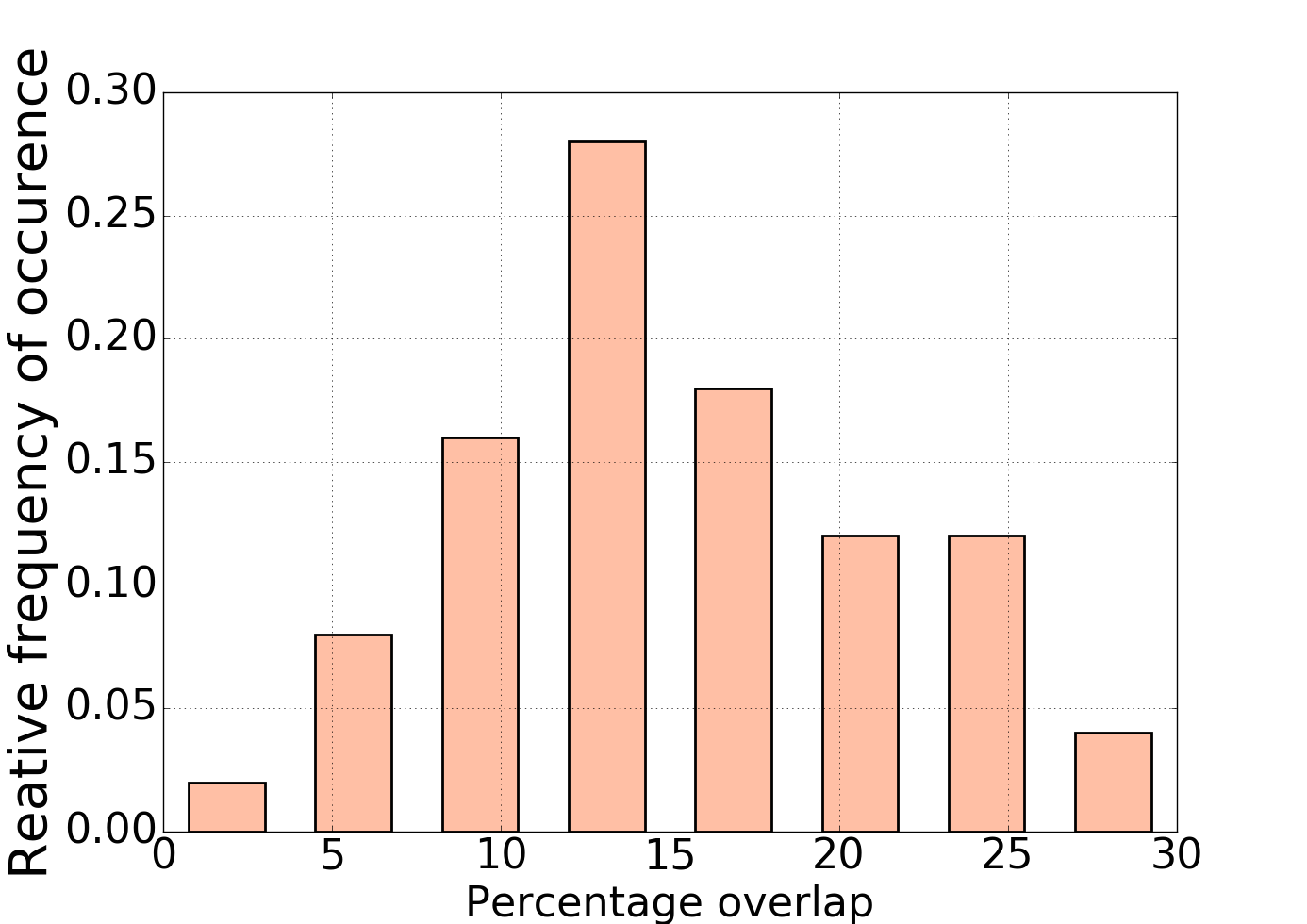}
  \caption{Distribution of the percentage overlap between the influencers identified by the IC model and the IC model with intrinsic activation in a Monte Carlo run.}
  \label{fig:hist}
\end{figure}

Next, for each of the Monte Carlo runs, we identify the percentage overlap between the sets of influential seeds identified by the IC model and our model with intrinsic activation. These results are shown in Figure \ref{fig:hist} for 50 Monte Carlo runs and for top 30 influential nodes. Note that while a small number of runs show nearly no overlap, more than 25\% of the runs show 20\% or more overlap. This is due to the fact that both the IC model and  our model with intrinsic activation favor nodes with large out-degrees. However, the IC model with intrinsic activation also requires that such nodes have a high enough $\alpha$ value to be influential along the lines of the discussions in sec. \ref{sec:local_influ}

Continuing our discussions that began in Section \ref{sec:local_influ} regarding maximizing the \textit{engagement}, we consider four different cases of assigning $\alpha$ values to the nodes.  In the first three cases, the random $\alpha$ values are drawn from three different intervals in a uniform manner. In the first case, all the $\alpha$ values are drawn according to the distribution $U[0,0.2]$. Meanwhile in the second case, they were drawn from $U[0.4,0.6]$ and finally in the third case, from the distribution $U[0.8,1.0]$. 

\begin{figure}
  \centering
  \includegraphics[width=10cm]{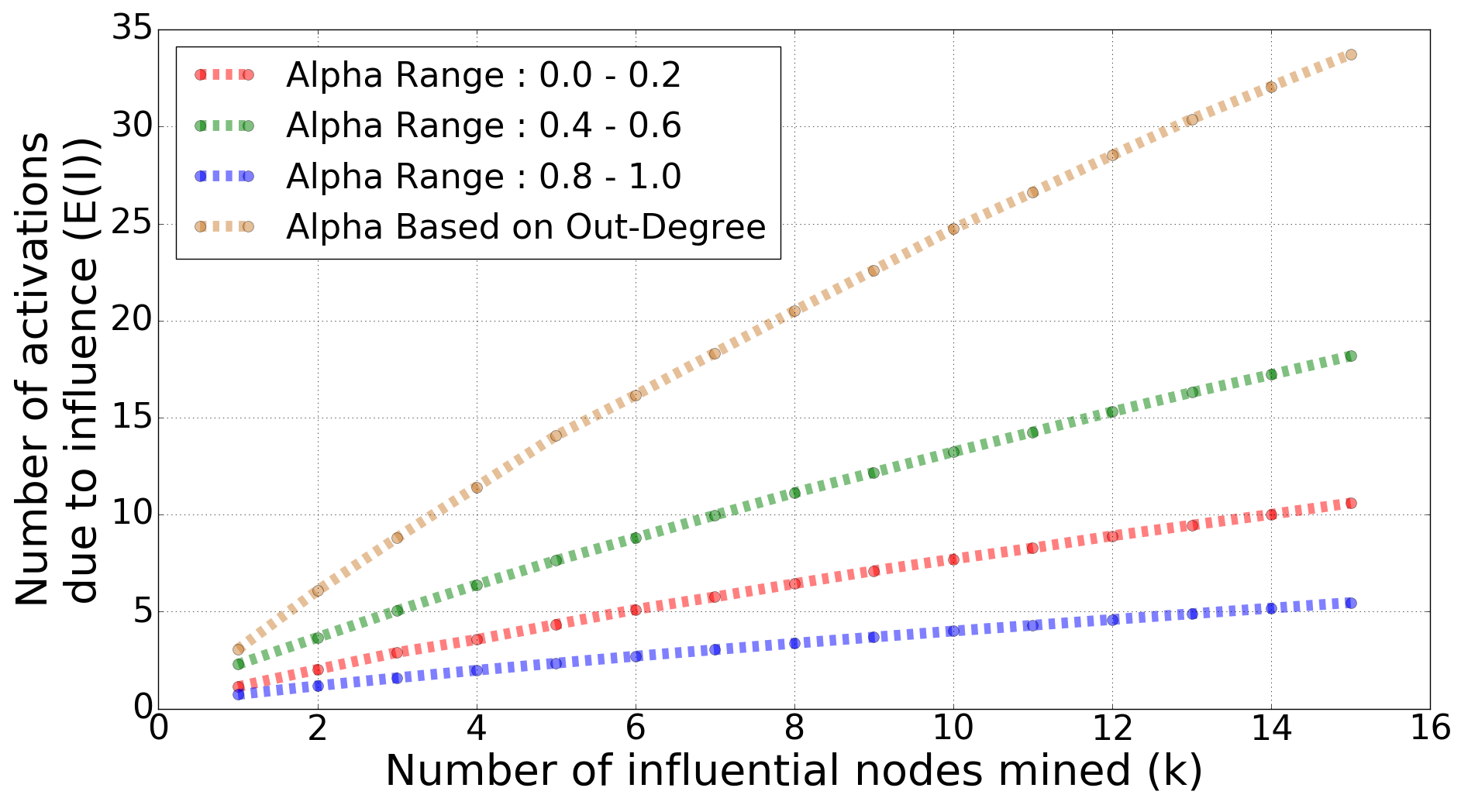}
  \caption{The mean activation curves for three different ranges of $\alpha$ assignments and the fourth with $\alpha$ values proportional to the node out-degree.}
  \label{fig:three_alphas}
\end{figure}

The mean number of influenced activations over three 30 run Monte Carlo analyses are plotted in Figure \ref{fig:three_alphas}. Note that each of the sub-problems corresponding to one realization of the $\alpha$ values for all the nodes, involved one run of Algorithm \ref{algo:fj-ic} with 1000 random samples ($n$). For the fourth case, the $\alpha$ values for the nodes are set deterministically in proportion to the  node out-degree values. Clearly, it can be seen that the cases where all of the $\alpha$ values  are either all small or all large fall short of the number of influenced activations for the case corresponding to the middle range of $\alpha$ values. The fourth case where the $\alpha$ values are proportional to the out-degree, far outperforms the rest. This observation is in line with the discussions presented in Section \ref{sec:local_influ}.

\section{A Centrality Metric Incorporating Intrinsic Activation}

In this final section, we examine the influencer mining on networks with intrinsic and influenced nodal activations from a slightly different perspective. By collecting the various probabilities together and recognizing the recursive nature of influence spread on a social network, we arrive at a generalized PageRank-type spectral influence measure that was first presented in \cite{sathanur2013physense}. As demonstrated in \cite{sathanur2013physense}, when considering activity on an OSN, this approach is a better measure of influence spread than a purely topological metric such as PageRank.   

For a given node $i$, from Figure \ref{fig:intro}, the total probability of activation $ p_{A}^T(i)$ can be written as

\begin{equation}
p_{A}^T(i) = \alpha(i) +\left(1-\prod_{j,(j,i) \in \mathcal{E}} \left(1-\beta(i)w_{ij}p_{A}^T(j)\right)\right).
\label{eqn:influVertex1}
\end{equation}
where $p_{A}^T(i)$ denotes the total probability of activation for node $i$ (intrinsic and influenced). The parameter $\alpha(i)$ denotes probability of node $i$ getting activated intrinsically. The quantity $\beta(i)w_{ij}$ denotes the probability of node $j$ activating node $i$ through influence as before.

 Equation \ref{eqn:influVertex1} summarizes the total activation probability for any given node in terms of the activation probabilities of the neighbors and whether the respective connecting edges are \textit{live} or not. For the IC model each activated node has a single shot probability of activating its neighbor and the activation along each edge is independent of the other edges. The second term in the large parentheses represents the probability of some neighboring node activating node $i$ through influence. When we write out Equation \ref{eqn:influVertex1} for all the nodes, we will be dealing with a large system of coupled non-linear equations, whose solution can be computationally expensive to obtain.  By retaining the leading-order terms, we get a linearized version of Equation \ref{eqn:influVertex1}. The goal of this approximation is to get the equation in a linear form so that we can use mature linear algebraic methods to quickly compute the set of influential nodes. This is valid to a large extent because, given the weighted-cascade version of the IC model that we are employing, the $w_{ij}$ values will be small and we can neglect the higher-order terms to linearize Equation \ref{eqn:influVertex1} as follows.

\begin{equation}
p_{A}^T(i) = \alpha(i) +\beta(i)\sum_{j,(j,i) \in \mathcal{E}} w_{ij}p_{A}^T(j).
\label{eqn:influVertex3}
\end{equation}

We set $\beta(i) = \left(1-\alpha(i)\right)$ as explained in Section \ref{sec:formulation} and extend Equation \ref{eqn:influVertex3} to the entire network with $N$ nodes to obtain a matrix-vector equation as follows:
\begin{equation}
\boldsymbol{p_{A}^T}= \boldsymbol{\alpha}\boldsymbol{\vec{1}}+\left(\left(\boldsymbol{I}-\boldsymbol{\alpha}\right)\boldsymbol{W}\right)\boldsymbol{p_{A}^T}.
\label{eqn:cmp_full_net1}
\end{equation}

In Equation~\ref{eqn:cmp_full_net1}, $ \boldsymbol{p_{A}^T}$  is a vector of size $N\times 1$, denoting respectively the total probability of activation for all the nodes on the network. $\boldsymbol{I}$ denotes the identity matrix of size $N\times N$. $\boldsymbol{\alpha}$ denotes the diagonal matrix with entries corresponding to the intrinsic activation probability for all the nodes on the network; $\boldsymbol{W}$ denotes the sparse, stochastic weight matrix with entries given by the weights $w_{ij}$ discussed earlier; and, $\boldsymbol{\vec{1}}$ is the all-ones vector of size  $N \times 1$.

We can then express the total activation probabilities as 
\begin{equation}
\boldsymbol{p_{A}^T}= \boldsymbol{\vec{1}}^T\boldsymbol{G}; G = \left(\boldsymbol{I} - \left(\boldsymbol{I}-\boldsymbol{\alpha})\boldsymbol{W}\right)\right)^{-1}\boldsymbol{\alpha},
\label{eqn:cmp_full_net2}
\end{equation}
where $\boldsymbol{\vec{1}}^T$ provides for the column-sum of $\boldsymbol{G}$. We also note that because the matrix $\boldsymbol{W}$ is a row stochastic matrix, the matrix $\boldsymbol{G}$ is also row stochastic. 

Consider the quantity $C_A(i)$, specific to node $i$ as defined below. 
\begin{equation}
C_A(i) = \left(\sum_{j=1,i \ne j}^{N}G_{ji}\right).
\label{eqn:centrality}
\end{equation}

$C_A(i)$ corresponds to the sum of the entries in column $i$ of $\boldsymbol{G}$ with the exception of the corresponding diagonal term and represents the expected number of hosts activated by node $i$ getting intrinsically activated and is a measure of influence.  In our experiments with the PBlogs and the LFR-1000 graphs, discussed in Section \ref{sec:larger_graphs}, the $\boldsymbol{\alpha}$ and $\boldsymbol{W}$ entries were randomized with entries drawn from the uniform distribution over $[0,1]$, and $\boldsymbol{W}$  was converted to a row-stochastic matrix. We then compare the sets of top-$k$ influencers identified by both the methods on two larger graphs in our dataset. The comparison is carried out with respect to two measures: 1) Jaccard similarity and 2) Rank Biased Overlap (RBO). RBO considers ordering with higher weights given to matches that happen at the top \cite{webber2010similarity}. These results are presented in Table \ref{table:comparison} where we see good agreement between the sets of influential nodes obtained by both methods. 

The behavior of the Jaccard index is not necessarily monotone as a larger number of influencers are considered. As we move away from the top influencing nodes (increasing $k$), we encounter many nodes that are of a similar influence. Since the centrality-based method is an approximation, the relative positions can change a lot and it is easily possible that going from k=30 to k=50, we may not get a proportional increase in the overlap between the two sets. Hence it can result in non-monotone behavior. The RBO based comparison can also exhibit a similar non-monotone behavior. However, this measure is known to be stable because of the weighting by the rank. The same is observed in our experimental results as well.

Thus, the proposed centrality metric, which includes the intrinsic activation mechanism, represents a computationally more viable alternative to the full-scale influence maximization framework. It retains the essence of the model and the influential nodes can be mined by solving a linear system involving a sparse matrix. 

\begin{table}[!htb]
\begin{center}
 \begin{tabular}{|c|c|c|c|c|c|} 
 \hline
Correlation type & Input & $k=10$ & $k=20$ & $k=30$ & $k=50$ \\ 
 \hline\hline  
Jaccard & PBlogs & 0.538 &  0.818  & 0.875 & 0.818  \\ 
 \hline 
RBO & PBlogs & 0.817 &  0.846  & 0.851 & 0.868  \\ 
 \hline
Jaccard & LFR1000 & 0.818 &  0.905  & 0.765 & 0.818  \\ 
 \hline 
RBO & LFR1000 & 0.979 &  0.963  & 0.947 & 0.937  \\ 
 \hline 
\end{tabular}
 \caption{Correlations, two ways, between the proposed approaches for the two inputs PBlogs and LFR1000 for different sizes of seed sets 
 (10, 20, 30, and 50). Closer the metric to one the better. }
 \label{table:comparison}
\end{center}
\end{table}

\section{Conclusions}

In this work, we introduce the notion of nodes in a complex network getting activated by two mechanisms: intrinsically and through influence as commonly
observed in online social networks. Using a modified version of the influence maximization algorithm and working with a suitable influence spread objective function, we show how it is possible to identify influential users whose intrinsic content spreads maximally through influence. We also sketch a short proof on the submodularity of the modified influence function, allowing for approximation guarantees on the algorithm. We utilized several synthetic and real-world datasets to examine various aspects of the proposed activation model. We also explain why some assignments of the intrinsic activation probability ($\alpha$ values) to the various nodes can result in much higher activations than other assignments, which is also demonstrated on a Twitter graph. We finally derive a novel centrality metric from the activation model that can provide for a computationally faster and accurate method to identify influential users on a social network where activations can be intrinsic or influenced. 

Building on this work, we are exploring multiple facets of this problem in our ongoing research including the exploration of how a social network can be successful in the long run by balancing the two modes of activation discussed here.  To achieve this objective, we are considering development of variants of other interaction models with intrinsic activation as well extending the notion of intrinsic activation to more fine-grained user behavior. We are also extending these methods to other complex systems, such as for attack modeling in cyber networks. 

 \section*{Acknowledgement}
This research was supported in part by the High Performance Data Analytics Program (HPDA) and in part by the Control of Complex Systems Initiative (CCSI) at the Pacific Northwest National Laboratory (PNNL). HPDA is a collaboration led by Pacific Northwest National Laboratory (PNNL) with partners Mississippi State University, University of Washington, and Georgia Institute of Technology. CCSI is a Laboratory Directed Research and Development (LDRD) program at the PNNL. PNNL is operated by Battelle for the U.S. Department of Energy under Contract DE-AC05-76RL01830.


\begin{thebibliography}{99.}
\providecommand{\natexlab}[1]{#1}
\providecommand{\url}[1]{{#1}}
\providecommand{\urlprefix}{URL }
\expandafter\ifx\csname urlstyle\endcsname\relax
  \providecommand{\doi}[1]{DOI~\discretionary{}{}{}#1}\else
  \providecommand{\doi}{DOI~\discretionary{}{}{}\begingroup
  \urlstyle{rm}\Url}\fi
\providecommand{\eprint}[2][]{\url{#2}}

\bibitem[{Twitter(2016)}]{TwitterAPI}
The Twitter Public API. \url{https://dev.twitter.com/rest/public}

\bibitem[{Adamic and Glance(2005)}]{adamic2005political}
Adamic LA, Glance N (2005) The political blogosphere and the 2004 us election:
  divided they blog. In: Proceedings of the 3rd international workshop on Link
  discovery, ACM, pp 36--43

\bibitem[{Bastian et~al(2009)Bastian, Heymann, and Jacomy}]{ICWSM09154}
Bastian M, Heymann S, Jacomy M (2009) Gephi: An open source software for
  exploring and manipulating networks

\bibitem[{Borgs et~al(2014)Borgs, Brautbar, Chayes, and Lucier}]{Borgs2014}
Borgs C, Brautbar M, Chayes J, Lucier B (2014) Maximizing social influence in
  nearly optimal time. In: Proceedings of the Twenty-Fifth Annual ACM-SIAM
  Symposium on Discrete Algorithms, SIAM, SODA '14, pp 946--957

\bibitem[{Chen et~al(2011)Chen, Collins, Cummings, Ke, Liu, Rincon, Sun, Wang,
  Wei, and Yuan}]{chen2011influence}
Chen W, Collins A, Cummings R, Ke T, Liu Z, Rincon D, Sun X, Wang Y, Wei W,
  Yuan Y (2011) Influence maximization in social networks when negative
  opinions may emerge and propagate. In: SIAM Data Mining, pp 379--390

\bibitem[{Cohen et~al(2014)Cohen, Delling, Pajor, and Werneck}]{Cohen2014}
Cohen E, Delling D, Pajor T, Werneck RF (2014) Sketch-based influence
  maximization and computation: Scaling up with guarantees. In: International
  Conference on Conference on Information and Knowledge Management, ACM, New
  York, NY, USA, CIKM '14, pp 629--638

\bibitem[{Farajtabar et~al(2014)Farajtabar, Du, Gomez-Rodriguez, Valera, Zha,
  and Song}]{farajtabar2014shaping}
Farajtabar M, Du N, Gomez-Rodriguez M, Valera I, Zha H, Song L (2014) Shaping
  social activity by incentivizing users. In: Advances in neural information
  processing systems, pp 2474--2482

\bibitem[{Friedkin and Johnsen(1999)}]{friedkin1999social}
Friedkin NE, Johnsen EC (1999) Social influence networks and opinion change.
  Advances in group processes 16(1):1--29

\bibitem[{Gionis et~al(2013)Gionis, Terzi, and Tsaparas}]{gionis2013opinion}
Gionis A, Terzi E, Tsaparas P (2013) Opinion maximization in social networks.
  In: SIAM Data Mining Conference, SIAM, pp 387--395

\bibitem[{Halappanavar et~al(2016)Halappanavar, Sathanur, and
  Nandi}]{Halappa:2016}
Halappanavar M, Sathanur A, Nandi A (2016) Accelerating the mining of
  influential nodes in complex networks through community detection. In:
  Proceedings of the 13th {ACM} International Conference on Computing
  Frontiers, CF'16, Como, Italy, May 16-18, 2016, pp~--

\bibitem[{Kempe et~al(2003)Kempe, Kleinberg, and Tardos}]{Kempe:2003}
Kempe D, Kleinberg J, Tardos E (2003) Maximizing the spread of influence
  through a social network. In: Proceedings of ACM SIGKDD, ACM, New York, NY,
  USA, pp 137--146, \doi{10.1145/956750.956769}

\bibitem[{Kempe et~al(2005)Kempe, Kleinberg, and Tardos}]{kempe2005influential}
Kempe D, Kleinberg J, Tardos {\'E} (2005) Influential nodes in a diffusion
  model for social networks. In: Automata, languages and programming, Springer,
  pp 1127--1138

\bibitem[{Krause and Golovin(2012)}]{krause2012submodular}
Krause A, Golovin D (2012) Submodular function maximization. Tractability:
  Practical Approaches to Hard Problems 3(19):8

\bibitem[{Lancichinetti and Fortunato(2009)}]{lancichinetti2009benchmarks}
Lancichinetti A, Fortunato S (2009) Benchmarks for testing community detection
  algorithms on directed and weighted graphs with overlapping communities.
  Physical Review E 80(1):016,118

\bibitem[{Leskovec et~al(2007)Leskovec, Krause, Guestrin, Faloutsos,
  VanBriesen, and Glance}]{leskovec2007cost}
Leskovec J, Krause A, Guestrin C, Faloutsos C, VanBriesen J, Glance N (2007)
  Cost-effective outbreak detection in networks. In: Proceedings of ACM SIGKDD,
  ACM, pp 420--429

\bibitem[{Li et~al(2013)Li, Chen, Wang, and Zhang}]{li2013influence}
Li Y, Chen W, Wang Y, Zhang ZL (2013) Influence diffusion dynamics and
  influence maximization in social networks with friend and foe relationships.
  In: Proceedings of the sixth ACM international conference on Web search and
  data mining, ACM, pp 657--666

\bibitem[{Mirzasoleiman et~al(2015)Mirzasoleiman, Badanidiyuru, Karbasi,
  Vondrak, and Krause}]{mirzasoleiman2015lazier}
Mirzasoleiman B, Badanidiyuru A, Karbasi A, Vondrak J, Krause A (2015) Lazier
  than lazy greedy. In: Twenty-Ninth AAAI Conference on Artificial Intelligence

\bibitem[{Myers et~al(2012)Myers, Zhu, and Leskovec}]{myers2012information}
Myers SA, Zhu C, Leskovec J (2012) Information diffusion and external influence
  in networks. In: ACM SIGKDD, ACM, pp 33--41

\bibitem[{Quach and Wendt(2016)}]{Quach2016}
Quach TT, Wendt JD (2016) A diffusion model for maximizing influence spread in
  large networks. In: Social Informatics: 8th International Conference, SocInfo
  2016, Bellevue, WA, USA, November 11-14, 2016, Proceedings, Part I, Springer
  International Publishing, pp 110--124

\bibitem[{Saito et~al(2008)Saito, Nakano, and Kimura}]{saito2008prediction}
Saito K, Nakano R, Kimura M (2008) Prediction of information diffusion
  probabilities for independent cascade model. In: International Conference on
  Knowledge-Based and Intelligent Information and Engineering Systems,
  Springer, pp 67--75

\bibitem[{Sathanur and Halappanavar(2016)}]{Sathanur2016}
Sathanur AV, Halappanavar M (2016) Influence maximization on complex networks
  with intrinsic nodal activation. In: Social Informatics: 8th International
  Conference, SocInfo 2016, Bellevue, WA, USA, November 11-14, 2016,
  Proceedings, Part II, Springer International Publishing

\bibitem[{Sathanur et~al(2013)Sathanur, Jandhyala, and
  Xing}]{sathanur2013physense}
Sathanur AV, Jandhyala V, Xing C (2013) Physense: Scalable sociological
  interaction models for influence estimation on online social networks. In:
  IEEE International Conference on Intelligence and Security Informatics, IEEE,
  pp 358--363

\bibitem[{Srivastava et~al(2014)Srivastava, Chelmis, and
  Prasanna}]{srivastava2014influence}
Srivastava A, Chelmis C, Prasanna VK (2014) Influence in social networks: A
  unified model? In: Advances in Social Networks Analysis and Mining (ASONAM),
  2014 IEEE/ACM International Conference on, IEEE, pp 451--454

\bibitem[{Webber et~al(2010)Webber, Moffat, and Zobel}]{webber2010similarity}
Webber W, Moffat A, Zobel J (2010) A similarity measure for indefinite
  rankings. ACM Transactions on Information Systems (TOIS) 28(4):20

\bibitem[{Zhang et~al(2014)Zhang, Mishra, and Thai}]{zhang2014recent}
Zhang H, Mishra S, Thai M (2014) Recent advances in information diffusion and
  influence maximization in complex social networks. Opportunistic Mobile
  Social Networks p~37

\end{thebibliography}

\end{document}